\documentclass[]{jfm}

\usepackage{graphicx}
\usepackage{newtxtext}
\usepackage{newtxmath}
\usepackage{natbib}
\usepackage{hyperref}
\hypersetup{
    colorlinks = true,
    urlcolor   = blue,
    citecolor  = black,
}

\newcommand{\RomanNumeralCaps}[1]
\linenumbers

\usepackage[amssymb]{SIunits}

\usepackage{color}

\title{Settling of chiral particles in a turbulent flow}

\author{Mees M. Flapper\aff{1}
\corresp{\email{m.m.flapper@utwente.nl}},
  John E. Sader\aff{2},
  Detlef Lohse\aff{1,3}
 \and Sander G. Huisman\aff{1}
  \corresp{\email{s.g.huisman@utwente.nl}}}

\affiliation{\aff{1}Physics of Fluids Department and Max Planck Center for Complex Fluid Dynamics, J. M. Burgers Centre for Fluid Dynamics, University of Twente, P.O. Box 217, 7522NB Enschede, The Netherlands
\aff{2}Graduate Aerospace Laboratories \& Department of Applied Physics and Materials Science, California Institute of Technology, Pasadena,CA 91125, USA
\aff{3}Max Planck Institute for Dynamics and Self-Organization, Am Faßberg 17, 37077, Göttingen, Germany}

\begin{document}
\maketitle

\begin{abstract}
Chiral particles are experimentally investigated while settling in water with various turbulence intensity levels. The locations and orientations of the particles are tracked over time, allowing the close investigation of the particles' settling dynamics. The generated turbulent flow is measured using laser Doppler anemometry (LDA), and the turbulence strength varies between experiments in the range $0 \leq \text{Re}_\lambda \leq 250$. Starting with quiescent particle settling, the chiral particle's orientation dynamics are studied, revealing a preferred alignment and a strong translation-rotation coupling. The particle chirality determines the preferred rotation direction, though the alignment and translation-rotation coupling gradually vanish with increasing turbulence. We identify multiple settling modes for the chiral particles, which are characterised by the evolution of the rotation angles. Finally, a theoretical model assuming a simplified chiral particle in Stokes flow clarifies the emergence of each settling mode.
\end{abstract}

\begin{keywords}
Authors should not enter keywords on the manuscript, as these must be chosen by the author during the online submission process and will then be added during the typesetting process (see \href{https://www.cambridge.org/core/journals/journal-of-fluid-mechanics/information/list-of-keywords}{Keyword PDF} for the full list).  Other classifications will be added at the same time.
\end{keywords}

\section{Introduction}
\label{sec:Introduction}

Prediction and control of particles in flows are important in a wide range of applications, in both natural and industrial contexts. For example, the climate and computational weather predictions of clouds containing ice hydrometeors are affected by the variation in shape and orientation of the hydrometeors \citep{Duncan2018}. Other applications include flow predictions for the dispersion of pollen \citep{Sabban2011,Roy2023} or the transport of floating plastic on the ocean surface \citep{vanSebille2020,Salmon2023}. In an industrial context, particle-laden flows have applications in pharmaceutical processes \citep{Erni2009}, and in recycling processes \citep{Pongstabodee2008,Chan2021}. Particles in these flows are generally non-spherical, which lead to a large range of complicated linear and angular velocity dynamics \citep{Voth2017}. The non-spherical shape of these particles introduces orientation-dependence of forces and torques on the suspended particles \citep{Marchioli2025}. Besides the shape of the particles, the turbulence of the particle-laden flow provides rich and complex dynamics: the parameter space is spanned by the particle-to-fluid density ratio, particle size compared to the smallest turbulence scales, and the solid-phase volume fraction \citep{Brandt2022}. Already on small scales, for laminar flow the dynamics of falling non-spherical particles in still water is extremely rich and counter-intuitive \citep{Joshi2025}.

In the context of studying ice hydrometeors, disks and perforated disks have been used to reproduce the settling dynamics of porous ice crystals and plates in a laboratory setting \citep{Tinklenberg2025}. The settling of millimetre-sized disks in quiescent air shows a bimodal distribution in falling modes, with disks either falling stably and aligned to the horizontal plane, or tumbling. However, the falling modes are non-trivially related to the particle size, with certain disk sizes displaying almost exclusively the tumbling falling mode \citep{Tinklenberg2023}. Introducing turbulence in this settling disk experiment significantly affects the settling dynamics for larger disks: the fall speed decreases for increasing disk size and turbulence strength, while the turbulence also causes the disks to tumble, leading to a more randomised orientation distribution. This being said, smaller disks (of sub-millimetre diameter) tend to fall aligned to the horizontal plane, regardless of the turbulence strength \citep{Tinklenberg2024}. Changing the disk shape by adding perforations affects the settling dynamics, where perforated disks are stabilised and tumble less frequently. In addition, the turbulence increases the drag coefficient for perforated disks through cross-flow-induced drag \citep{Tinklenberg2025}.

These results show that the shape, size, and turbulence all significantly affect the settling dynamics of disks in non-trivial ways. Going beyond disks, rods and fibres show much different dynamics in turbulence compared to disks. Here, rods tend to spin around their symmetry axis more than they tumble, whereas disks tumble more than they spin, due to the different alignment with the fluid vorticity \citep{Byron2015}. The resulting mean square rotation rates for disks are much higher compared to spheres or rods \citep{Parsa2012}. The mean square rotation rate of the rods is also found to scale as $\propto l^{-4/3}$ in the inertial range of turbulence, where $l$ is the rod length \citep{Parsa2014}. A similar scaling is found for spheroids, for which the mean square angular velocity is shown to also scale as $l^{-4/3}$, but here $l$ is given by the volume-equivalent diameter \citep{Jiang2022}. Once again, this illustrates the vast parameter space affecting the dynamics of anisotropic particles in turbulence, and emphasises that particle dynamics are affected by the particle's shape, size, and the ambient turbulence.\\

This study focuses on an even wider class than symmetrical particles, namely on chiral particles, which break mirror symmetry. On the very smallest scales, chiral molecules have applications in chemistry \citep{Roos2015}, and many biological systems are often intrinsically chiral \citep{Chan2021}. An example of a chiral particle is a maple seed, which autorotates as it descends, increasing its lift, thereby increasing the distance it can travel \citep{Lentink2009,Sohn2022}. This translation-rotation coupling is further illustrated in recent work by \cite{Huseby2025}, who use helical ribbons with a strong translation-rotation coupling. Settling these ribbons in a quiescent fluid at very low Reynolds numbers results in quasi-periodic angular dynamics and complex spatial trajectories. Here the helical ribbon length and the initial orientation strongly affect the dynamics and resulting trajectory.

Chiral particles in turbulence have been investigated by \cite{Kramel2016}, showing that a chiral dipole (a rod with two chiral ends) has a preferential rotation in homogeneous isotropic turbulence, resulting from an alignment of the particle with the fluid strain. More recently, \cite{Piumini2024} computationally investigated heavy chiral particles $\left( 2 \leq \rho_p/\rho_f \leq 10 \right)$ in quiescent fluid and homogeneous isotropic turbulence. In a quiescent fluid, the particles display a translation-rotation coupling, where the rotation direction is determined by the particle chirality. As the turbulence is increased, the chirality-induced mean rotation decreases, such that the chirality does not matter at high Reynolds numbers. Other studies have pointed out that the rotation direction of a particle is not solely determined by the inherent chirality of the particle \citep{Moreno2024}.

\begin{figure}
     \centering
     \includegraphics[width=\linewidth]{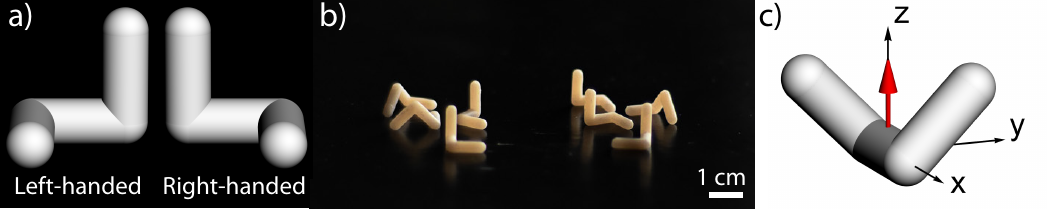}
     \caption{a) A 3D visualisation of a chiral enantiomorph, displaying a left-handed and right-handed particle. b) A photograph of the 3D-printed chiral particles used in experiments. c) A 3D visualisation of a chiral particle in its reference orientation, including axes, and a pointing vector shown by the red arrow.}
     \label{fig:Particles_photo}
 \end{figure}

This study experimentally and theoretically studies the settling dynamics of heavy chiral particles in a quiescent fluid and in turbulence, employing the same chiral particles as used by \cite{Piumini2024}. The particle geometries used here are illustrated in Figure \ref{fig:Particles_photo}a), and a photograph of the particles is shown in Figure \ref{fig:Particles_photo}b). The reference orientation of the particles is shown in Figure \ref{fig:Particles_photo}c), with the red arrow indicating the pointing vector.
 The particles consist of three equally large rods, all being mutually orthogonal. This allows for two different particle chiralities, giving left-handed and right-handed particles. This specific particle shape was chosen due its simplicity, while still breaking mirror symmetry.
 
 We also compare and contrast the findings of chiral particles in turbulence with previously found computational results. We aim to characterise the dynamics of settling chiral particles in a quiescent flow, and illustrate any differences in dynamics caused by the particle chirality. By introducing and increasing turbulence in the settling dynamics, we highlight how the particle dynamics change as a result of the turbulence strength. Based on these results, we can conclude whether any difference in dynamics between the particle chiralities vanish at high turbulence intensity, as seen computationally.
 
 The paper is organised as follows: section \ref{sec:Experimental setup} shows the experimental setup used to perform the experiments, and characterises the turbulent flow used for the settling experiments. Furthermore, this section uses a simple model to obtain a first estimate of the particle's settling velocity and angular velocity. Section \ref{sec:Results} describes the results and compares findings from the present study to previous computational work. Section \ref{sec:theoretical clarification} presents a simple theoretical model for particle settling that incorporates the effect of turbulence with stochastic forcing. We use this model to clarify our experimental findings. Finally, section \ref{sec:Conclusion and outlook} provides concluding remarks and an outlook.

\section{Experimental setup}\label{sec:Experimental setup}
\begin{figure}
    \centering
    \includegraphics[width=0.9\linewidth]{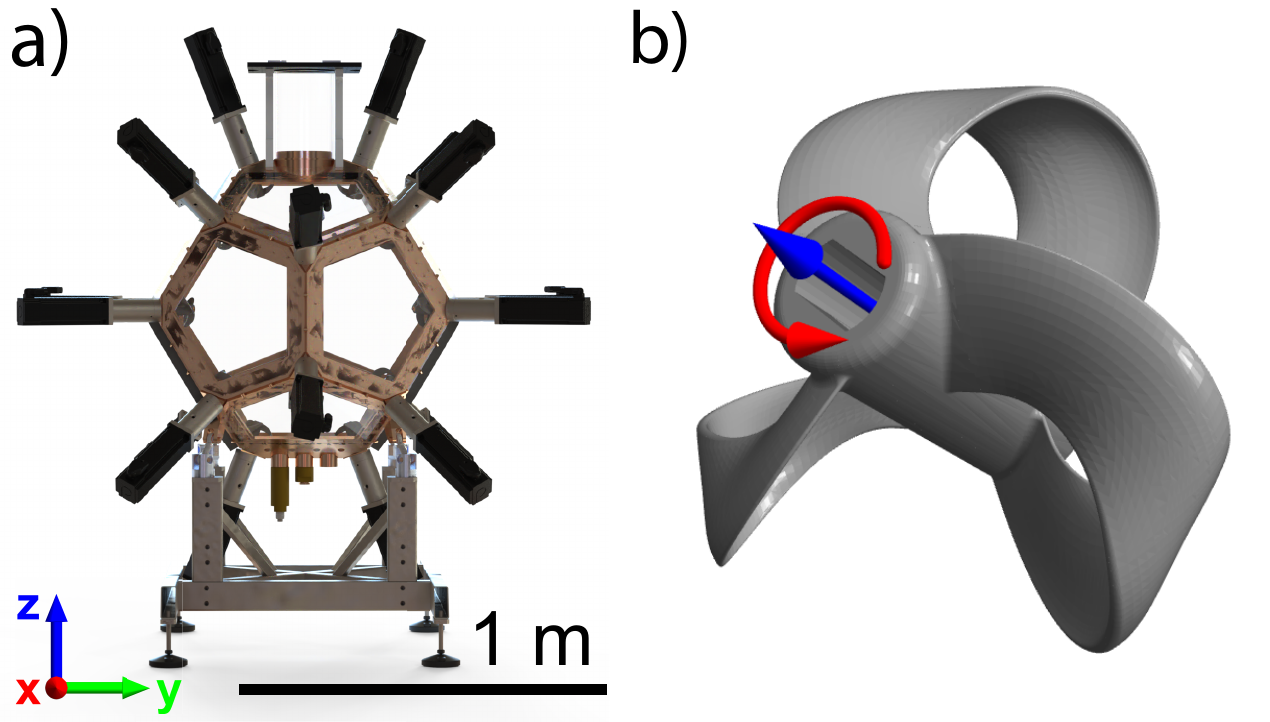}
    \caption{a) A render of the Dodecahedron setup, including a definition of the lab coordinate axes, where the origin is at the centre of the setup. b) A 3D drawing of one of the toroidal propellers used inside the Dodecahedron. The blue arrow indicates the axis of rotation, the red arrow shows the positive rotation direction.}
    \label{fig:Dodecahedron setup}
\end{figure}

The particles used in the settling experiments are 3D-printed using a Formlabs Form 3 printer. Figure \ref{fig:Particles_photo}b) shows the chiral particles, which have a characteristic size of $l = $ \unit{11.3}{\milli \meter} (the length of one straight line segment of the particle, termed a `rod' of the particle), or an end-to-end Euclidean length of approximately \unit{17}{\milli \meter}, and a density of $\rho =$ \unit{1.30 \cdot 10^3}{kg/\meter^3}. The diameter of each rod is $d =$ \unit{3.2}{\milli \meter}. We let the particles settle in our dodecahedron setup (henceforth simply termed the `Dodecahedron'), which is shown in figure \ref{fig:Dodecahedron setup}a) and is based on the LEM (Lagrangian Exploration Module) setup described by \cite{Zimmermann2010}.
The Dodecahedron has edge lengths of \unit{30}{\centi \meter}, and a total volume of approximately $V =$ \unit{210}{\liter}. For these experiments, the Dodecahedron is filled with water. The setup is equipped with 20 engines (Beckhoff AM8033), which have a maximum power of \unit{1}{\kilo \watt} each. Each engine rotates a 3D-printed toroidal propeller, shown in figure \ref{fig:Dodecahedron setup}b). The propeller's shape causes the fluid displacement to differ substantially depending on the direction in which it rotates. This is further shown in section \ref{subsec:Flow measurement}, where we characterise the flow in the Dodecahedron using laser Doppler anemometry.

To perform the settling experiments, the chiral particles were released in water one-by-one at the top of the Dodecahedron. We released the particles by hand, with a random initial orientation, approximately \unit{55}{\centi \meter} above the centre of the setup, such that the particles reach their terminal velocity in the measurement volume. The particles were tracked in the centre of the setup, where we used a single Photron Nova S16, and two Photron Mini AX-200 high-speed cameras (all \unit{1024}{px} $\times$ \unit{1024}{px}) to record the settling particles. Two cameras are mounted to the large windows on the front of the setup, while the third camera is mounted on the smaller window on top. The cameras image a measurement volume of approximately \unit{10}{\centi \meter} $\times$ \unit{19}{\centi \meter} $\times$ \unit{15}{\centi \meter}, with a resolution of approximately \unit{160}{\micro \meter / px} for the cameras on the front windows, while the top camera has a resolution of approximately \unit{175}{\micro \meter / px}. All cameras record using a frame rate of \unit{250}{fps}.

\begin{figure}
    \centering
    \includegraphics[width=0.8\linewidth]{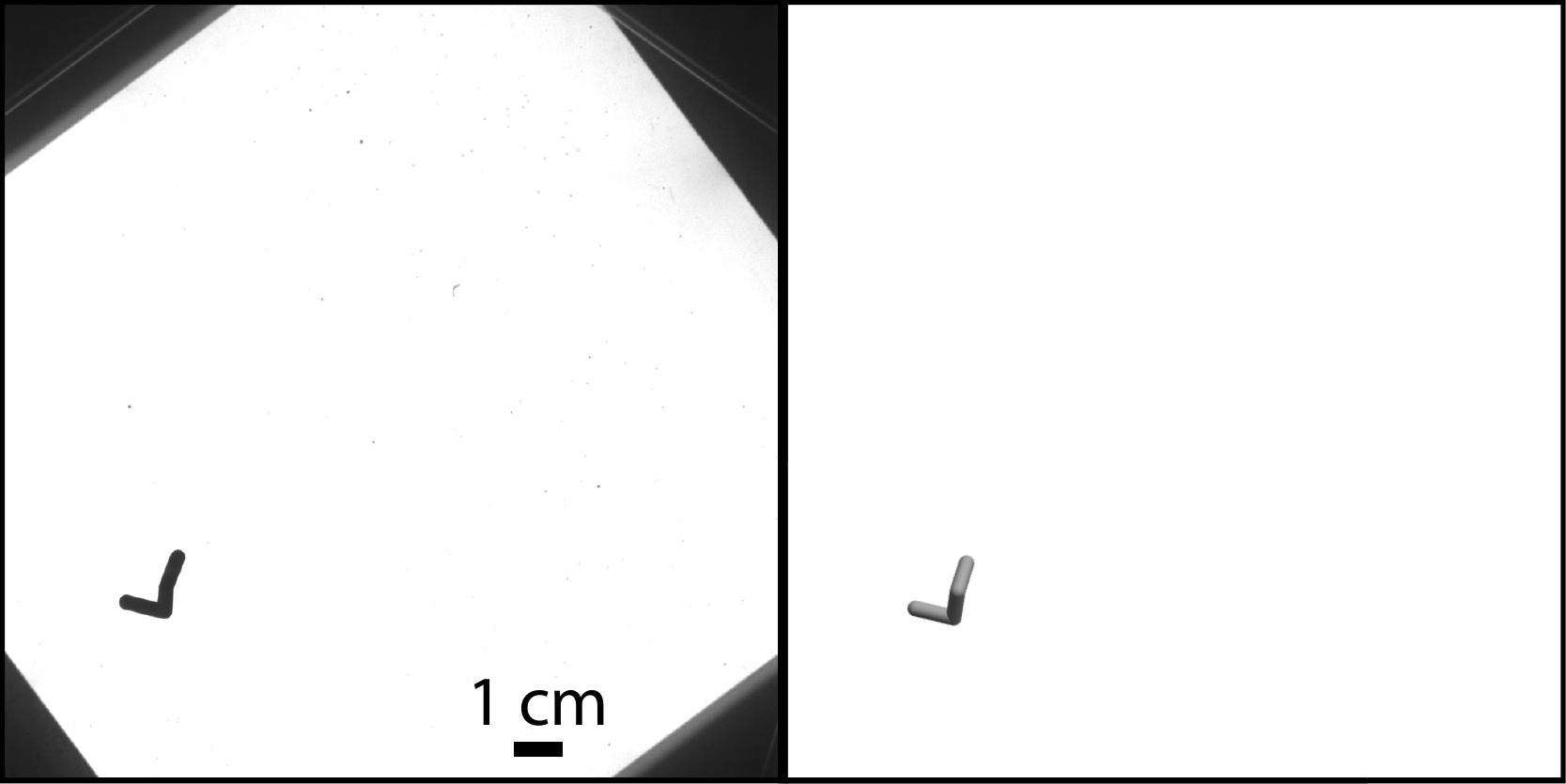}
    \caption{A raw image of a recorded chiral particle (left panel), and its corresponding 3D reconstruction (right panel). The method for reconstructing the particle is described in \cite{Flapper2025}.}
    \label{fig:Chiral_Particle_reconstruction_frame}
\end{figure}

The particles are tracked using the orientation tracking algorithm described in \cite{Flapper2025}, specifically designed for tracking the orientation of anisotropic particles, where the algorithm can distinguish the particle chirality. This method creates a synthetic particle with a known orientation, and calculates the synthetic projections seen by the cameras. The synthetic particle projections are compared to the experimental images, and the synthetic particle's orientation is varied to minimise the difference between experimental images and synthetic projections. The resulting orientation of the synthetic particle then gives the orientation of the experimentally imaged particle. A 3D chiral particle is then reconstructed by using the found orientation. An example of a recorded image and the reconstructed particle is shown in Figure \ref{fig:Chiral_Particle_reconstruction_frame}. An animation of a recorded video alongside the reconstructed particles is shown in the supplementary material.

To investigate the effect of turbulence on the settling dynamics, the Reynolds number was varied by rotating the propellers at different rotation rates: the propellers were rotated at \unit{0}{Hz}, \unit{-0.25}{Hz}, \unit{-0.5}{Hz}, \unit{-1}{Hz}, \unit{-2}{Hz}, and \unit{-4}{Hz}, with all propellers rotating in the same direction at the same rotation rate. The highest absolute rotation rate of \unit{-4}{Hz} was chosen due to it being the highest absolute rotation rate where the particles still sank. The flow generated by this setup was characterised using laser Doppler anemometry (LDA) to find the resulting flow parameters such as the velocity fluctuations, isotropy, the energy dissipation rate, and the Taylor--Reynolds number. The findings and results from these measurements are detailed below.

\subsection{Turbulent flow characterisation} \label{subsec:Flow measurement}
\begin{figure}
    \centering
    \includegraphics[width=0.7\linewidth]{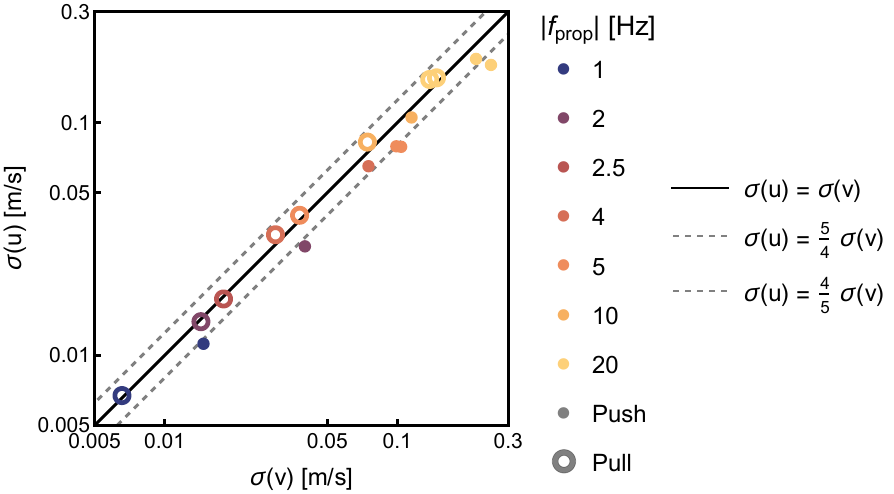}
    \caption{Standard deviation in the horizontal and vertical velocity components plotted against one another for numerous propeller rotation frequencies. Added lines indicate ratios between the velocity standard deviations.}
    \label{fig:Isotropy_plot}
\end{figure}

Using laser Doppler anemometry, the $y$-component and $z$-component of the velocity ($u$ and $v$ respectively) are measured in the centre of the Dodecahedron. We record velocity time series for absolute rotation rates between \unit{1}{Hz} to \unit{20}{Hz} in both rotation directions. The lower bound of \unit{1}{Hz} was used since lower propeller rotation speed resulted in too low velocity fluctuations to be measured reliably. Each measurement collects data for \unit{30}{minutes}, in which the propellers rotate at a constant frequency. The collected velocity data is used to calculate the standard deviations in velocity $\sigma(u)$ and $\sigma(v)$ for each rotation frequency, the ratio of which indicates the isotropy of the flow. Figure \ref{fig:Isotropy_plot} shows this plot, illustrating the standard deviations in both velocity components. Again, we note that the difference between the positive and negative rotation frequency is caused by the propeller shape: the negative rotation frequencies `push' the fluid towards the centre of the Dodecahedron, while the positive rotation frequencies `pull' the fluid from the centre. For the negative rotation frequencies (`pushing' the fluid), the ratio of velocity standard deviations is around $\sigma(u)/\sigma(v) = 4/5$, whereas the positive rotation frequencies are all in the range $ 0.93 \leq \sigma(u)/\sigma(v) \leq 1.11$. This indicates better isotropy when pulling the fluid, compared to pushing. Looking at the values of the standard deviation in both velocity components, we observe that the absolute values are consistently higher for pushing compared to pulling, indicating a higher turbulence strength for pushing.

Characterising the strength of the turbulence is done by calculating the energy dissipation rate $\epsilon$, computed from the second order structure function. Employing the Kolmogorov hypotheses for large $r$ in the inertial range $(\eta < r < L)$, the relation between the longitudinal structure function and the energy dissipation rate is given by (assuming no intermittency corrections)
\begin{equation}
    D_{LL}(r) = C_2 (\epsilon r)^{2/3}.
\end{equation}
Here $D_{LL}$ is the longitudinal structure function, $r$ is the distance between two points, and $C_2 = 2.0$ is a universal constant \citep{Pope}. From this equation, the energy dissipation rate can be calculated as
\begin{equation}
    \epsilon = \frac{1}{r} \left(\frac{D_{LL}(r)}{C_2} \right)^{3/2},
\end{equation}
which can be estimated by finding a plateau of the compensated structure function in the inertial range. Since the LDA probes a single position over a long period of time, the separation in time is converted to a separation in space by using the measured velocity and the difference in arrival time \citep{Buchhave2017}.
\begin{figure}
    \centering
    \includegraphics[width=0.8\linewidth]{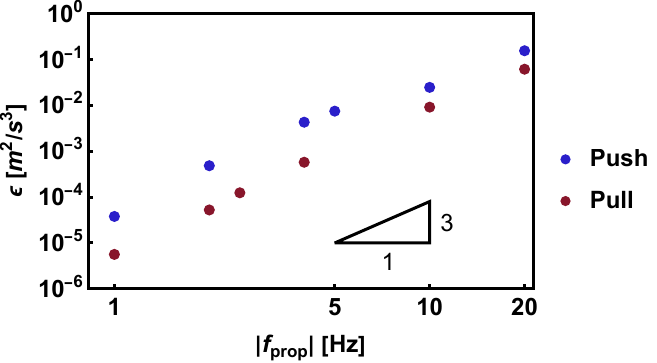}
    \caption{The energy dissipation rate $\epsilon$ as a function of the propeller rotation frequency. The black triangle shows the expected cubic scaling between the energy dissipation rate and the frequency.}
    \label{fig:Epsilon_plot}
\end{figure}
Figure \ref{fig:Epsilon_plot} shows the calculated values of the energy dissipation rate for the measured propeller frequencies on a logarithmic plot. The colour of the points show the rotation direction of the propeller. The black triangle shows the expected cubic scaling between the energy dissipation rate and the propeller rotation frequency, where the measurements follow the expected scaling well. \\
For the settling experiments reported below, propeller rotation rates from \unit{0}{Hz} to \unit{-4}{Hz} are used, where we choose to use negative rotation rates due to the higher energy dissipation rate, despite higher anisotropy. The most important flow parameters at the various rotation rates are displayed in table \ref{tab:Flow parameters}. The reported values are based on the measured longitudinal structure function, and the resulting values of $\epsilon$, where the data for the rotation rates lower than \unit{1}{Hz} are extrapolated from the measured data.
\begin{table}
    \centering
    \begin{tabular}{cccccc}
    \hline
       $\text{f}_{\text{prop}}$ [Hz] & $\sigma(v)$ [m/s] & $\epsilon$ [$\text{m}^2/\text{s}^3$] & Re$_\lambda$ & Re$_\text{p}$ & Stk\\
    \hline
        $-0.25$ & $3.2 \cdot 10^{-3}$ & $4.3 \cdot 10^{-7}$ & 29 & 36 & 1.1\\
        $-0.5$ & $6.7 \cdot 10^{-3}$ & $4.3 \cdot 10^{-6}$& 49 & 76 & 2.3\\
        $-1$ & $1.4 \cdot 10^{-2}$ & $3.8 \cdot 10^{-5}$& 79 & 159 & 4.8\\
        $-2$ & $3.0 \cdot 10^{-2}$ & $4.8 \cdot 10^{-4}$& 153 & 341 & 10\\
        $-4$ & $6.6 \cdot 10^{-2}$ & $4.3 \cdot 10^{-3}$& 250 & 749 & 23\\
        \hline
    \end{tabular}
    \caption{Turbulent flow parameters.}
    \label{tab:Flow parameters}
\end{table}
The ambient turbulence is characterised by the Taylor--Reynolds number, given by 
\begin{equation}
    \text{Re}_{\lambda} = \frac{\sigma(v)\lambda}{\nu} \text{,}
    \label{eq:Rel}
\end{equation}
where $\lambda = \sigma(v)\sqrt{15 \nu/\epsilon}$ is the Taylor length. The particle Reynolds number $\text{Re}_\text{p}$ is computed as 
\begin{equation}
    \text{Re}_\text{p} = \frac{\sigma(v) l}{\nu},
    \label{eq:Rep}
\end{equation}
whereas the Stokes number is defined as 
\begin{equation}
    \text{Stk} = \frac{t_0 \ \sigma(v)}{l} \text{,} 
\end{equation}
with $t_0 = \rho_p d_{\text{eq}}^2/(18 \mu)$. Here, as above, $\rho_p$ is the particle density and $l$ the particle length scale, $d_{\text{eq}}$ is the equivalent diameter of a sphere of the same volume as a chiral particle, and $\mu$ is the dynamic viscosity of water. Besides the flow conditions listed in table \ref{tab:Flow parameters}, we also performed an experiment in quiescent water. For the quiescent settling case, the most relevant parameter is the Galileo number, given by 
\begin{equation}
    \text{Ga} = \frac{l \sqrt{g l \left(\frac{\rho_p}{\rho_f} - 1 \right)}}{\nu},
\end{equation}
resulting in a value of $\text{Ga} = 2.1\cdot10^3$.

Settling measurements were performed in quiescent water and in each of the turbulent flows reported in table \ref{tab:Flow parameters}. For each measurement, approximately 90 particles of each chirality were tracked while settling.

\section{Results} \label{sec:Results}
\begin{figure}
    \centering
    \includegraphics[width=\linewidth]{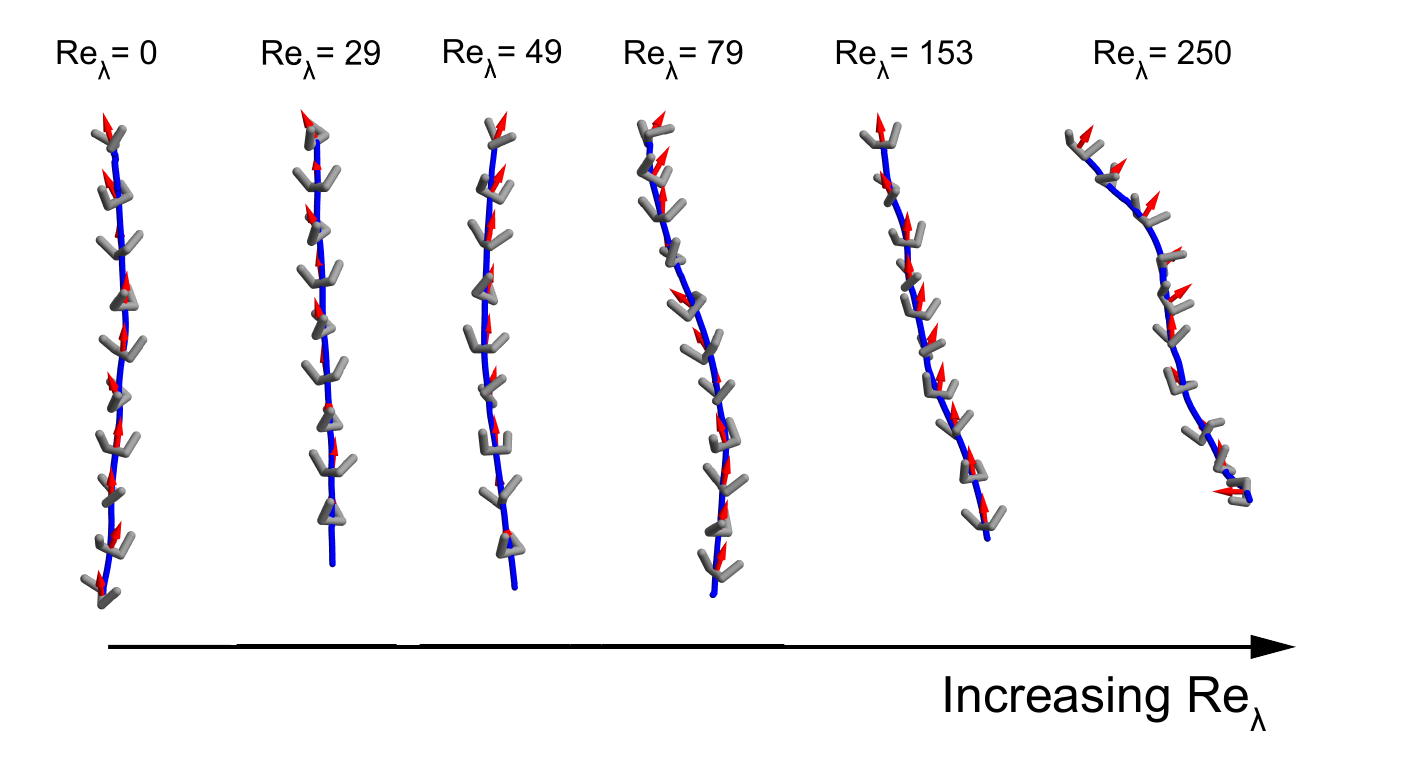}
    \caption{Snapshots of settling chiral particles at the different turbulence levels. All shown particles are left-handed, with snapshots spaced \unit{0.14}{s} apart. The blue line shows the centre of mass trajectory, the red arrow indicates the pointing vector.}
    \label{fig:Snapshots_settling_chiral}
\end{figure}

We track the settling chiral particles, measuring their position and orientation over time. The linear and angular velocities of the particles are obtained through time differentiation of these results. This procedure is performed across a range of Reynolds numbers. An overview of the general results are illustrated in Figure \ref{fig:Snapshots_settling_chiral}, showing snapshots of settling chiral particles (all left-handed) in flows of the various measured turbulence levels. The snapshots are all spaced by \unit{0.14}{s}, and the blue line shows the centre of mass trajectory, whereas the red arrow represents the pointing vector. An animated version of this figure is included in the supplementary material, which more clearly illustrates the particle dynamics.
The figure shows that the particle aligns with its pointing vector opposing the direction of gravity for small Taylor--Reynolds number, though the alignment weakens and vanishes as the turbulence increases. Similarly, we can see a clear preferential rotation around the vertical when studying the snapshots, which demonstrates a translation-rotation coupling. Again, this preferential rotation vanishes when increasing the Reynolds numbers. Finally, the horizontal movement of the particle is observed to increase with the turbulence intensity. Overall, we see the transition from a stable settling mode, with clear alignment and rotation, to a more irregular falling mode (tumbling) as the turbulence is increased. The same observations are made when examining right-handed particles, where only the rotation direction around the vertical is reversed. These observations are studied in more detail in the following sections.

\begin{figure}
    \centering
    \includegraphics[width=0.5\linewidth]{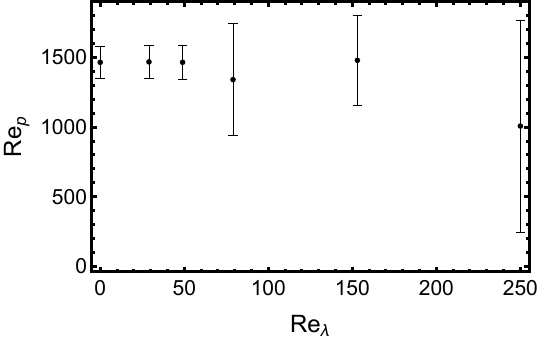}
    \caption{Particle Reynolds number based on the settling velocity as function of the Taylor--Reynolds number.}
    \label{fig:Mean settle vels}
\end{figure}

The mean settling velocity in the vertical direction is calculated per particle, and averaged per Taylor--Reynolds number, where each particle's velocity is weighted by the duration of the recorded trajectories.
The mean settling velocity is then nondimensionalised and thus expressed as particle Reynolds number $\text{Re}_\text{p}$, as defined in \eqref{eq:Rep}, using the characteristic rod length $l$ and the kinematic viscosity $\nu$. This particle Reynolds number is plotted against the Taylor--Reynolds number, $\text{Re}_\lambda$, in figure \ref{fig:Mean settle vels}.
The interval markers show a standard deviation in the average settling velocity (nondimensionalised into $\text{Re}_\text{p}$). The plot illustrates that the average settling velocity is not strongly affected at low Taylor--Reynolds numbers, but the variation in velocity increases with the Taylor--Reynolds number. Only for the highest Taylor--Reynolds number do we see a large decrease in average settling velocity. For this data point, the standard deviation is very large, which is due to the large velocity fluctuations with respect to the quiescent falling velocity. The overall finding from this plot is in agreement with previous studies, which found a reduction in settling velocity for heavy anisotropic particles as a result of turbulence \citep{Good2014,Chan2021,Tinklenberg2024}.\\

\begin{figure}
    \centering
    \includegraphics[width=\linewidth]{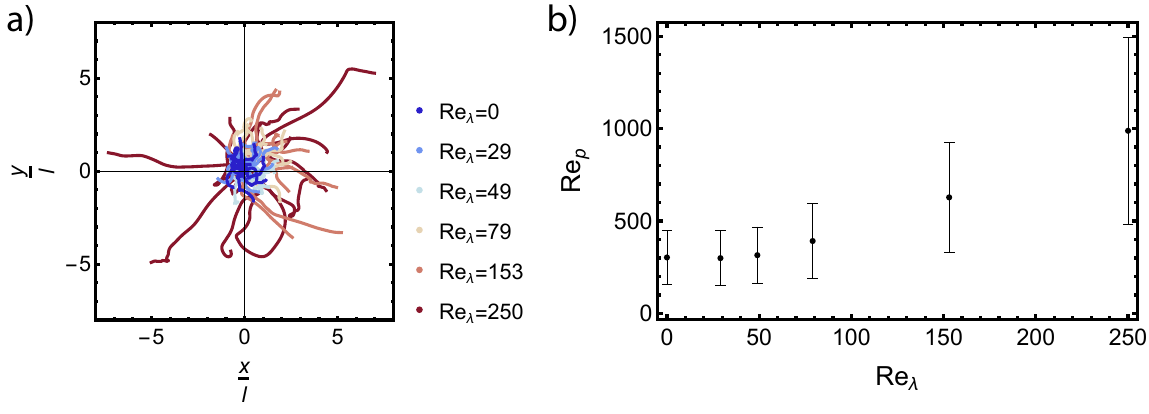}
    \caption{a) Top-view of the trajectories of a selection of the chiral particles, for all measured turbulence levels. The trajectories have been shifted to start at the origin, all shown trajectories are taken over the same time span. b) The particle Reynolds numbers based on the root mean square of the horizontal velocity. Error bars denote the standard deviation.}
    \label{fig:Trajectory_Replanar}
\end{figure}

The overall particle velocity is in the downward direction, though the particle velocity in the horizontal plane increases with turbulence. This is evident from studying the top-view of the particle trajectories, shown in Figure \ref{fig:Trajectory_Replanar}a).
This plot shows a number of projected trajectories for all measured turbulence levels, with all trajectories measured over the same time duration. The trajectories show an increase in traversed horizontal distance with increasing turbulence, which is further illustrated and quantified in Figure \ref{fig:Trajectory_Replanar}b). For this figure, the particle Reynolds number is computed using the root mean square of the horizontal velocity of the settling particles, where the error bars represent a standard deviation. The plot clearly illustrates an increase in the horizontal particle velocity and its variation as the turbulence intensity increases. The results so far highlight a decrease in settling velocity and an increase in horizontal velocity as the turbulence strength increases, i.e., due to velocity fluctuations in the base flow generated by the propellers. Besides these effects of the turbulence on the translation of the particles, our main interests are in the particle orientation dynamics and how these are affected by the imposed turbulence.

\subsection{Orientation data}
\begin{figure}
    \centering
    \includegraphics[width=0.8\linewidth]{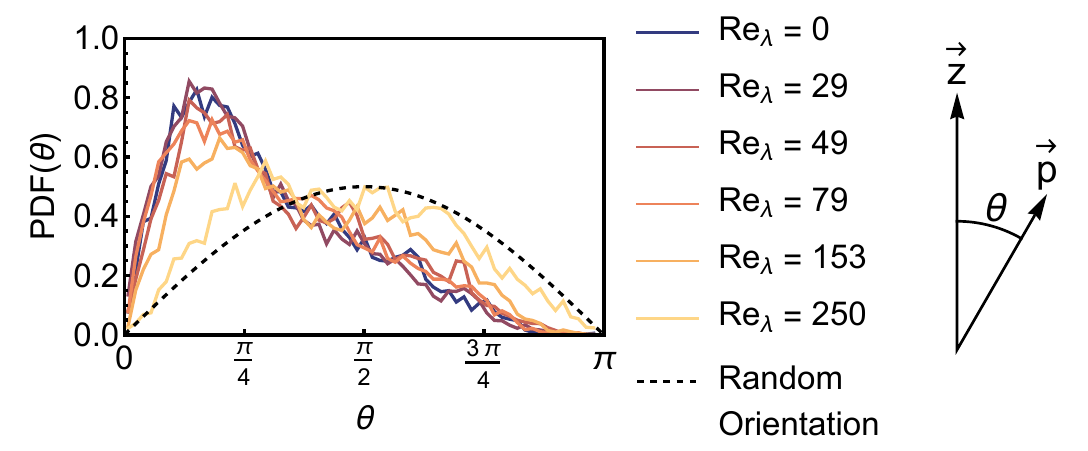}
    \caption{PDFs of the angle between the particle pointing vector and the vertical for all measured Reynolds numbers. The black dashed line indicates a random distribution. The graphic on the right shows the definition of the alignment angle.}
    \label{fig:upvecs_pdf}
\end{figure}

The reconstructed chiral particles allow us to closely study the orientation dynamics. First, we investigate whether the particles preferentially align as found by \cite{Piumini2024}, and how any preferential alignment is affected by the turbulence. Computing the particle's alignment is done by finding the pointing vector over time, where the pointing vector is chosen as shown in figure \ref{fig:Particles_photo}c). The angle between the pointing vector and the vertical is then computed to find the alignment between the particle and the direction of gravity. The collected results of the alignment angle are shown in figure \ref{fig:upvecs_pdf}, which shows the probability density function (PDF) of the alignment angle for the different Reynolds numbers. The definition of the alignment angle is shown by the graphic on the right. Here we note that the particle's rotation around the pointing vector is immaterial. The dashed line in the figure illustrates the curve for a randomly distributed orientation for reference.
The PDFs of the alignment angle show a clear preferential alignment, as shown by the large deviation with respect to the random distribution: the pointing vector preferentially aligns opposite to the direction of gravity, as was also observed in numerical work by \cite{Piumini2024}. This preferential alignment is visible for the lower Reynolds numbers, and vanishes for the high Reynolds numbers. At the highest measured Reynolds numbers, the particle orientation is therefore no longer determined by the geometry of the particle, but rather by the imposed turbulent flow. This figure therefore illustrates a preferential alignment of the particle, which weakens and vanishes for increasing turbulence strength.

\subsection{Rotation data}
\begin{figure}
    \centering
    \includegraphics[width=\linewidth]{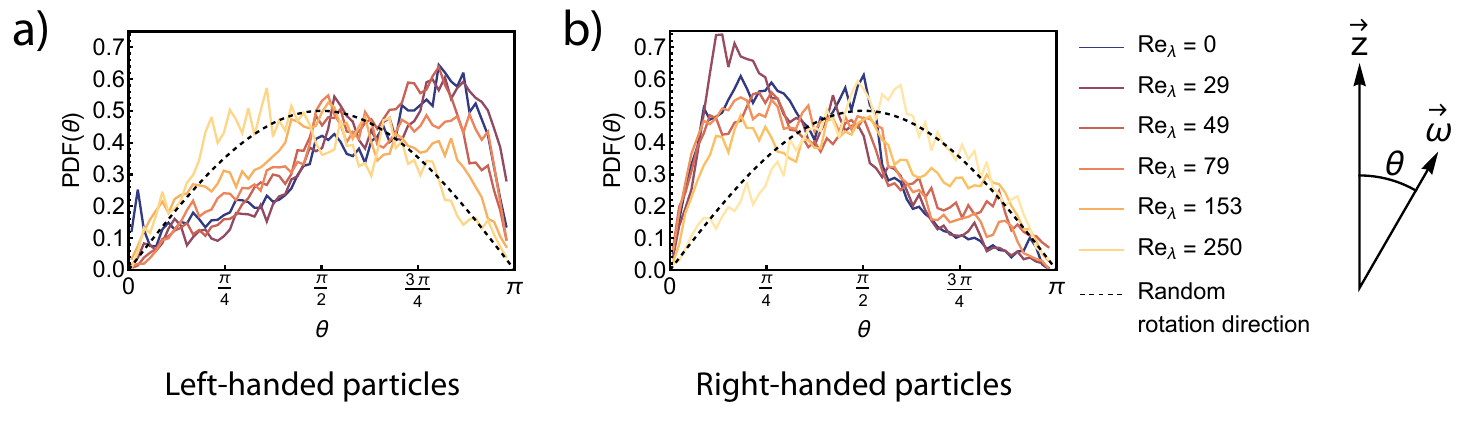}
    \caption{PDF of the angle between the rotation vector and the vertical for (a) left-handed particles and (b) right-handed particles.}
    \label{fig:PDF rotation vectors}
\end{figure}

Similar to the alignment of the particle's pointing vector, we find the alignment of the rotation vector with respect to the vertical. The PDF of the angle between the rotation vector with the vertical is shown in figure \ref{fig:PDF rotation vectors}, where the data is split between the left-handed and right-handed particles.
Figure \ref{fig:PDF rotation vectors}a) shows that the left-handed particles preferentially have their rotation vector pointing in the direction of gravity, opposing the particles' pointing vector. On the other hand, figure \ref{fig:PDF rotation vectors}b) illustrates that the right-handed particles have their rotation vector pointing opposite to the direction of gravity, aligning with the particles' pointing vector. Comparing these results emphasises that the particle chirality dictates the rotation direction during settling, highlighting the translation-rotation coupling for these particles. This experimental finding is consistent with the numerical observations by \cite{Piumini2024}, showing a difference in rotation direction for the left-handed and right-handed particles.

Comparing the rotation vector alignment across the measured Taylor--Reynolds numbers, $\text{Re}_\lambda$, we find that the preferential rotation becomes less pronounced as the Taylor--Reynolds number increases. For the highest Taylor--Reynolds numbers, no preferential rotation can be distinguished, and the effect of the particle chirality on its rotation statistics vanishes, which is in line with the previous numerical simulations by \cite{Piumini2024}.

\subsection{Multiple settling modes}
Besides the stable and tumbling settling mode illustrated in Figure \ref{fig:Snapshots_settling_chiral}, another settling mode is observed, though rarely, in experiments. In contrast, in the numerical work, only the previously shown stable settling mode at low Reynolds numbers and a tumbling motion at higher Reynolds numbers was found.

\begin{figure}
    \centering
    \includegraphics[width=\linewidth]{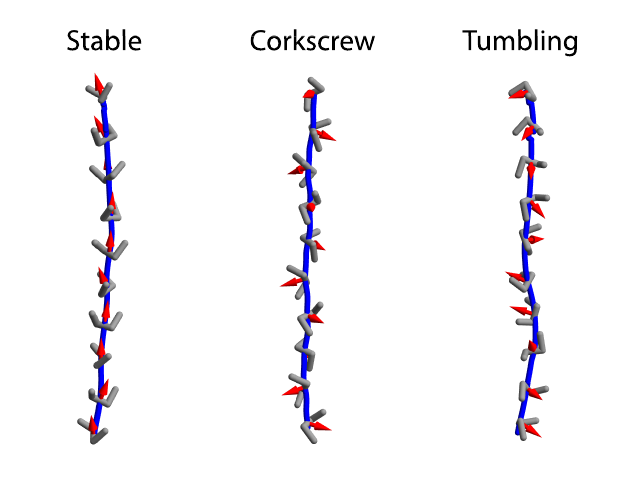}
    \caption{Snapshots of all observed falling modes of the chiral particles (all left-handed). The snapshots are separated by \unit{0.14}{s}, the blue line represents the centre of mass trajectory, and the red arrow illustrates the pointing vector. The snapshots are all taken from measurements at $\text{Re}_\lambda = 0$.}
    \label{fig:Falling_styles}
\end{figure}

We call this new mode the `corkscrew' mode. The corkscrew mode is illustrated alongside the stable and tumbling mode in Figure \ref{fig:Falling_styles}, showing snapshots of the different falling dynamics side-by-side. An animated version of this figure is added in the supplementary material, showing the dynamics of the different settling modes more clearly.
This figure shows the clear difference in orientation between the stable and corkscrew modes, even though both modes rotate around the vertical axis, and have a straight, vertical trajectory. In contrast, the tumbling mode does not show a clear alignment or rotation. We hypothesise that the corkscrew mode remains stable as it minimises the `frontal area' of the particle while settling, by aligning the bottom rod of the particle to the fluid. Since this settling mode is not observed in simulations, this hypothesis remains unverified.

\begin{figure}
    \centering
    \includegraphics[width=\linewidth]{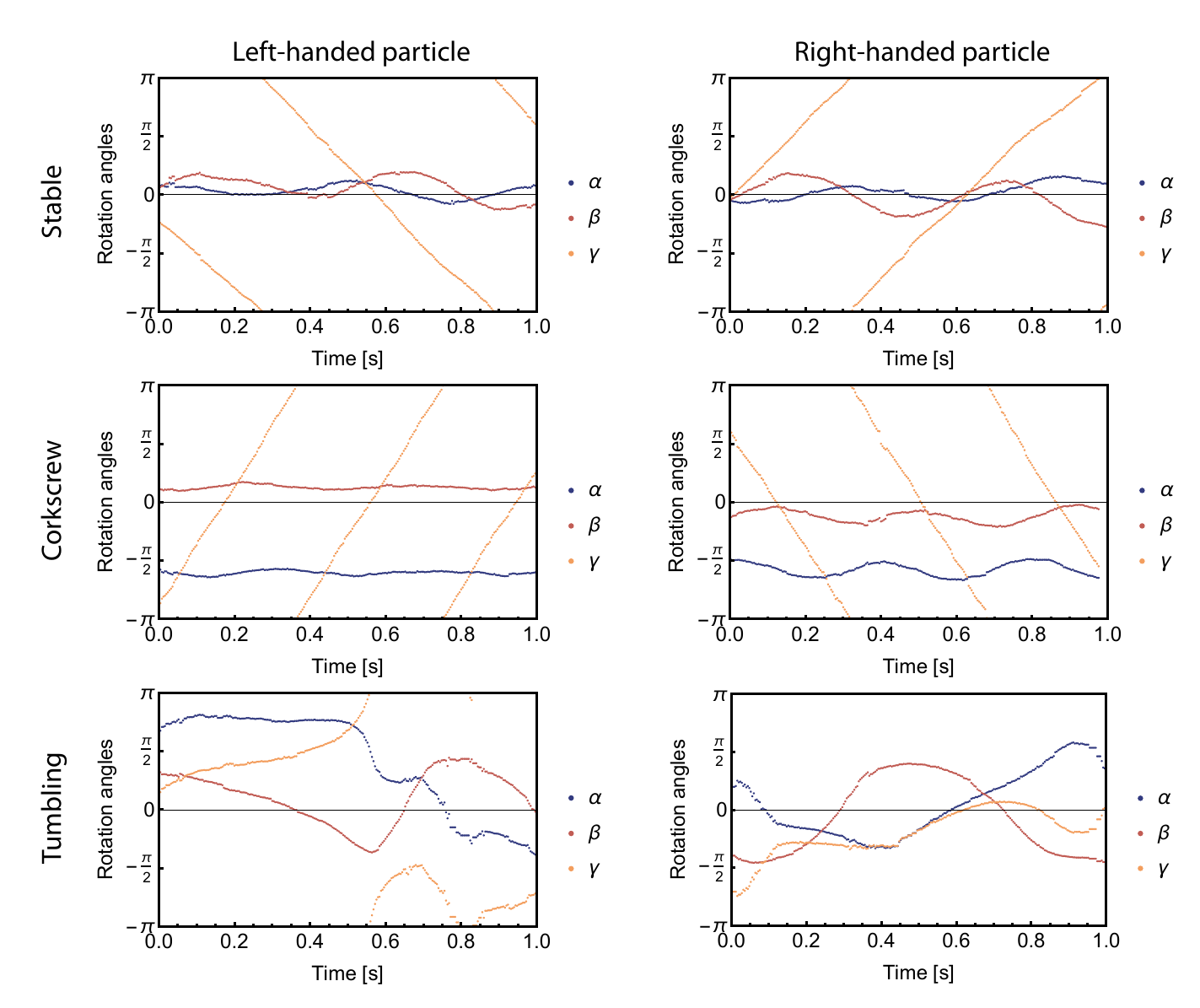}
    \caption{The rotation angles over time for all observed falling modes, and both particle chiralities. The rotations are with respect to the reference orientation, with the $x$, $y$, and $z-$axis as rotation axis (in this order). $\alpha$, $\beta$, and $\gamma$ denote the rotations around the $x$, $y$, and $z-$axis respectively.}
    \label{fig:Rotangs_fallstyles}
\end{figure}

These three experimentally found settling modes are studied in detail by examining the rotation angles of the particle over time for both particle chiralities, cf. Figure \ref{fig:Rotangs_fallstyles}.
The plots show the particle's rotation around the $x,y$, and $z$ axes with respect to the reference orientation shown in Figure \ref{fig:Particles_photo}c). Here $\alpha$ gives the rotation around the $x$-axis, $\beta$ around the $y$-axis, and $\gamma$ around the $z$-axis, with the order of rotations $x,y,z$. 
The stable settling mode is characterised, as noted previously, by the rotation around the vertical axis, as illustrated by the angle $\gamma$ decreasing monotonically for the left-handed particle. The angles $\alpha$ and $\beta$ oscillate around $0$, indicating that the pointing vector oscillates closely around the vertical, as seen in Figure \ref{fig:upvecs_pdf} for low Taylor--Reynolds numbers. On the other hand, the right-handed particle shows a similar profile, though $\gamma$ now increases monotonically, highlighting the differing rotation direction as a result of the particle chirality.

The `corkscrew' mode is remarkable due to the particle rotating around the $z$-axis in the opposite direction compared to the regular stable falling mode, as $\gamma$ now increases monotonically for the left-handed particle. We note that the rotation angles $\alpha$ and $\beta$ oscillate little, and are not centred around $0$. The angle $\alpha$ is close to $-\pi/2$, indicating that the pointing vector (approximately) lies in the $x,y$-plane. This is in contrast to the stable falling mode, where the pointing vector aligns opposite to gravity. Again, the right-handed particle shows similar rotation angles with a reversed rotation direction around the $z$-axis. A last point worth noting is that the angular velocity of the corkscrew is slightly higher compared to the stable settling mode: $\omega =$ \unit{10}{rad/s} for the corkscrew mode, compared to $\omega=$ \unit{7}{rad/s} for the stable settling chiral particles. Finally, the tumbling mode displays a more random orientation over time. The orientation angles show no clear pattern, nor periodicity.

\begin{figure}
    \centering
    \includegraphics[width=\linewidth]{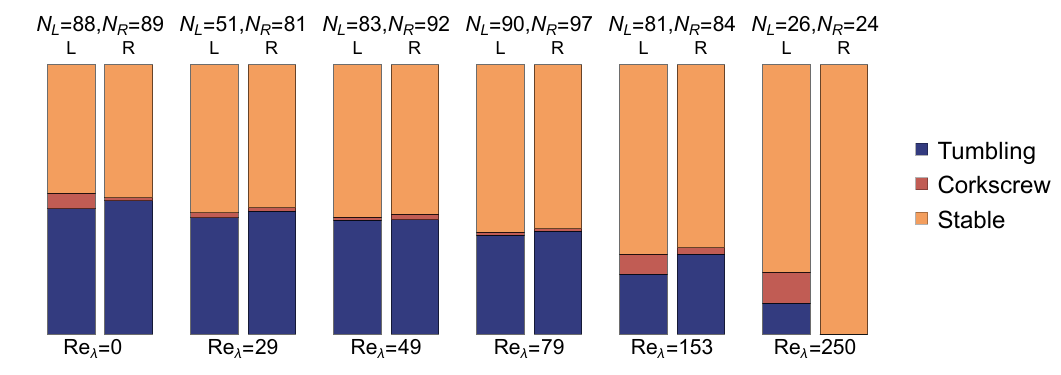}
    \caption{Percentage of occurrence for each falling modes, shown for all measured Taylor--Reynolds numbers, split between particle chirality. The numbers above the bar chart indicate the number of tracked left-handed and right-handed particles.}
    \label{fig:Tally-fallstyles}
\end{figure}

These different settling modes are tallied to evaluate their frequency of occurrence. Using a $k$-means clustering algorithm, each particle track is classified as stable, corkscrew, or tumbling. This classification is done based on 12 parameters per particle track: for each rotation angle we compute the mean, slope, linear fit residuals, and standard deviation of the residuals. The clustering is applied to particle tracks longer than $100$ frames, which we find to give a good indication of the settling mode. The clusters are checked manually and corrected where necessary: the size of the clusters indicate how frequently each settling mode occurs at each turbulence level. This is shown in Figure \ref{fig:Tally-fallstyles}, indicating how often each settling mode occurs in percentages. The data is split by particle chirality, and the numbers above the bars indicate how many particles are tracked for each Taylor--Reynolds number (split by particle chirality).
This affirms earlier findings, emphasising the common occurrence of the stable settling mode at lower turbulence levels, which becomes rarer as the turbulence increases. The fact that settling modes besides tumbling are still observed at large Taylor--Reynolds numbers indicates that preferential alignment and rotation can still occur at surprisingly high turbulence strengths. Another point we observe is that the corkscrew settling mode only occurs rarely: based on the rare occurrence, and the absence of this settling mode in numerical simulations, the cause for this settling mode cannot be readily identified from the obtained results.

The absence of the corkscrew settling mode in simulations calls for extra investigation: computationally, different initial orientations were explored, which were unable to produce the corkscrew mode. Experimentally, extra measurements were performed, which showed that the corkscrew mode is likely to occur when a bubble is attached to the end of a chiral particle, which leads to a distorted density distribution.

\section{Theoretical model} \label{sec:theoretical clarification}
\begin{figure}
    \centering
    \includegraphics[width=0.4\linewidth]{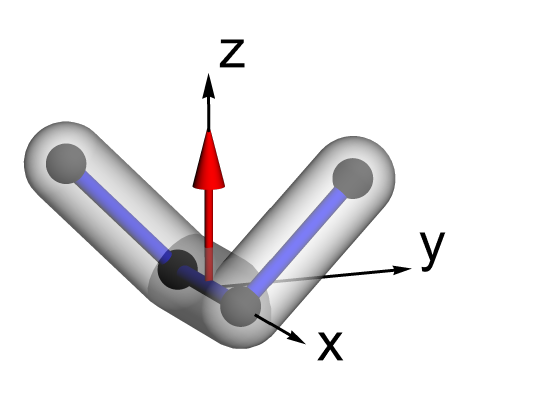}
    \caption{Simplified chiral particle consisting of spheres that are rigidly connected, visualised by the black spheres, with blue rods demonstrating the rigid connections. The semi-transparent tube illustrates the particle used in experiments. The red vector again illustrates the chosen pointing vector.}
    \label{fig:Chiral part theory}
\end{figure}

To clarify and further investigate the chiral particle's settling dynamics, we develop a model based on a simplified chiral particle consisting of four rigidly-connected spheres, as shown in Figure \ref{fig:Chiral part theory}. This replaces the true particle geometry with a minimal set of spheres that are chosen to capture the chiral particle’s symmetry and its dominant fluid-structure physics.
The settling dynamics of this particle, assuming a Stokes flow under gravity, are calculated
using standard hydrodynamics that employs the Oseen mobility tensor to account for the hydrodynamic coupling between the spheres. This enables the velocity of each sphere to be calculated provided spheres are well separated relative to their radii: in our case the separation length is unity, whereas the radii of the spheres is $1/10$. The force on each sphere is used to find the net force and torque experienced by the particle as a whole, which allow us to find its translational velocity, \textbf{U}, and its angular velocity, $\boldsymbol{\Omega}$. The particle position and orientation are then time-stepped by solving for \textbf{U} and $\boldsymbol{\Omega}$ at each time. The effect of the imposed turbulent flow is modelled through the inclusion of a stochastic force. The details of this theoretical model are provided in Appendix \ref{app: theoretical model}.

\begin{figure}
    \centering
    \includegraphics[width=0.5\linewidth]{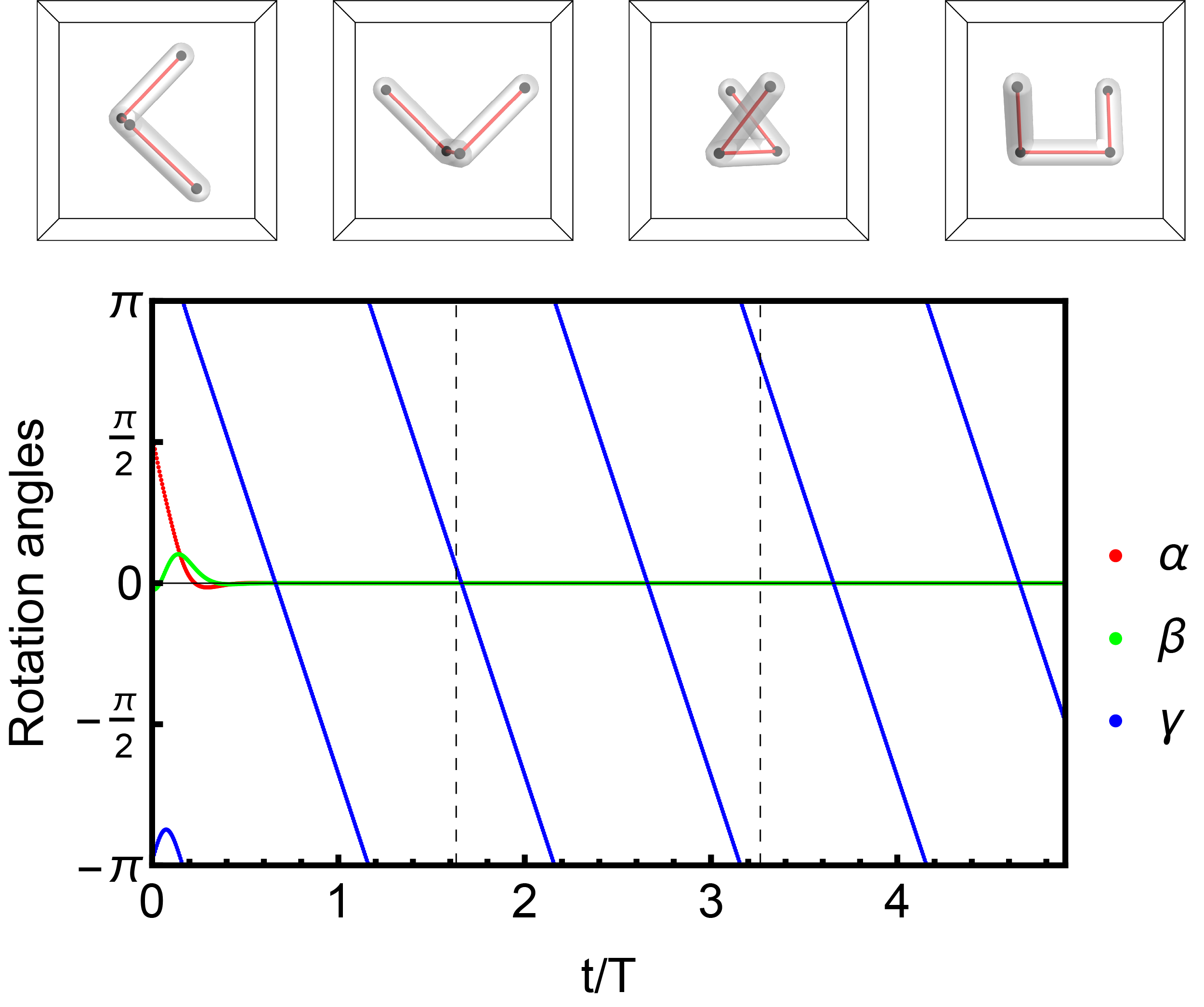}
    \caption{Theoretical model: Rotation angles for a left-handed chiral particle settling in Stokes flow. The snapshots illustrate the particle orientation at different time steps.}
    \label{fig:rotangs_stable_theory}
\end{figure}

\begin{figure}
    \centering
    \includegraphics[width=0.5\linewidth]{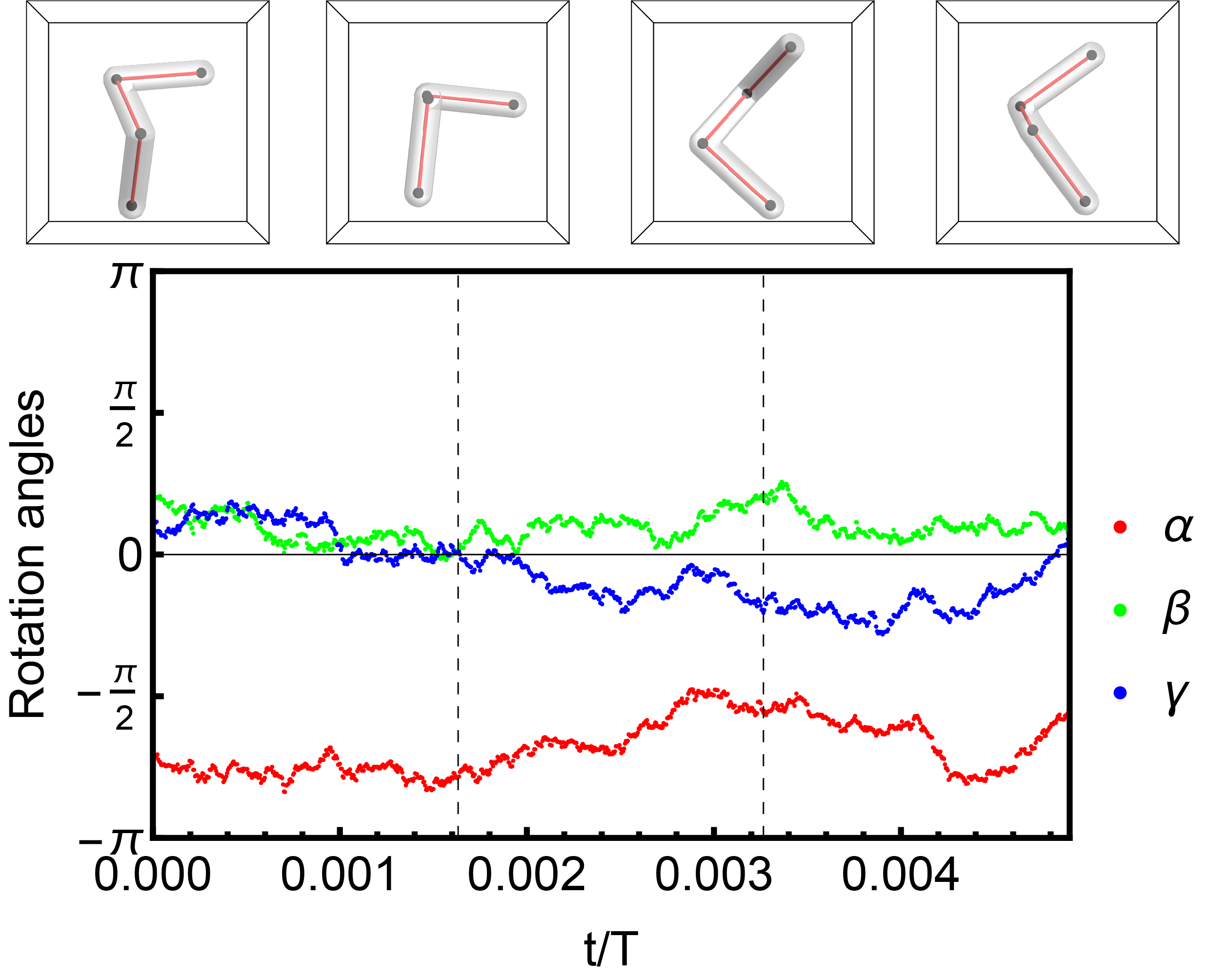}
    \caption{Theoretical model: Rotation angles for a left-handed chiral particle settling in Stokes flow, with a stochastic turbulent forcing. The stochastic force has half the magnitude of the force of gravity. The snapshots illustrate the particle orientation at different time steps.}
    \label{fig:rotangs_tumbling_theory}
\end{figure}

The dynamics of the settling chiral particle in the absence of turbulence, i.e., without any stochastic forcing, show a stable settling mode, which closely matches the experimental findings. Snapshots of the settling chiral particle along with the rotation angles over time are shown in Figure \ref{fig:rotangs_stable_theory} for the left particle chirality. 
The characteristic alignment and preferential rotation of the stable settling particle are observed, as previously found in experiments and simulations. In the supplementary material, an animation of a settling left-handed and right-handed chiral particle is shown, clearly displaying the preferential alignment and rotation. In the presence of turbulence, modelled by stochastic forcing of the particle (see above), we see a more irregular settling motion, similar to the tumbling mode in experiments. The rotation angles and snapshots for a tumbling particle are shown in Figure \ref{fig:rotangs_tumbling_theory}, where the magnitude of the stochastic forcing is half the magnitude of the force of gravity (that is to say, the standard deviation of the stochastic force is half the magnitude of the force of gravity). Again, an animation of a settling chiral particle with turbulent forcing is included in the supplementary material. These findings illustrate that the stable and tumbling settling modes are readily found from the theoretical model.

\begin{figure}
    \centering
    \includegraphics[width=\linewidth]{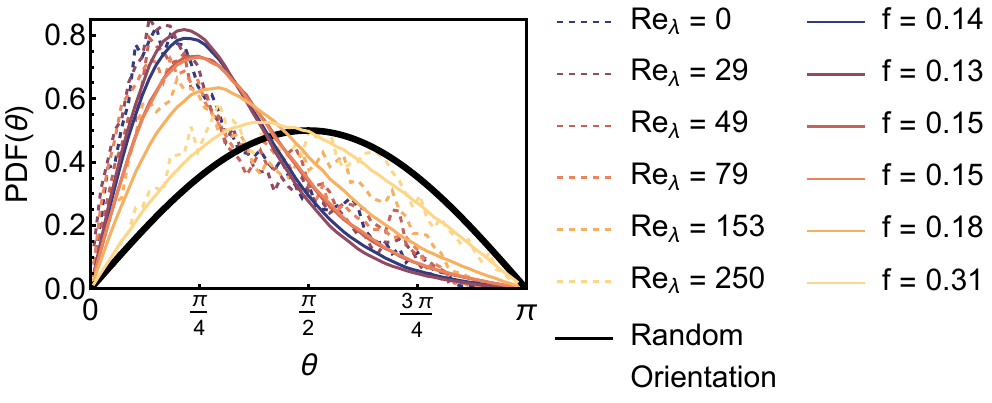}
    \caption{PDF of the angle between the pointing vector and the vertical for experiments (dashed lines), and theoretical model (solid lines). The value $f$ indicates the magnitude of the stochastic forcing as a fraction of the force of gravity.}
    \label{fig:upvecs_theory_pdf}
\end{figure}

In experiments with increasing level of turbulence, a transition from the stable settling mode to the tumbling mode was observed, as characterised by the particle orientation and rotation alignment, cf. Figure \ref{fig:upvecs_pdf} and \ref{fig:PDF rotation vectors}. We check whether a similar stable to tumbling transition occurs in the theoretical model by varying the strength of its stochastic forcing. The PDF for the angle between the vertical and the pointing vector is shown in Figure \ref{fig:upvecs_theory_pdf}. 
The magnitude of the stochastic force, $f$, is a fraction of the magnitude of the force of gravity: these values were chosen by fitting the PDFs of the theory to those of the experiments. Fitting the PDFs was done by minimising the sum of squared distances between the experimental and theoretical PDFs, where each fit uses a dataset of 5000 to 10000 settling chiral particles from the theoretical model.
The plot shows a closely matching trend between experiments and theory: this again confirms that the preferential alignment of the chiral particle occurs at low turbulence intensity, and vanishes with increasing turbulence.

\begin{figure}
    \centering
    \includegraphics[width=\linewidth]{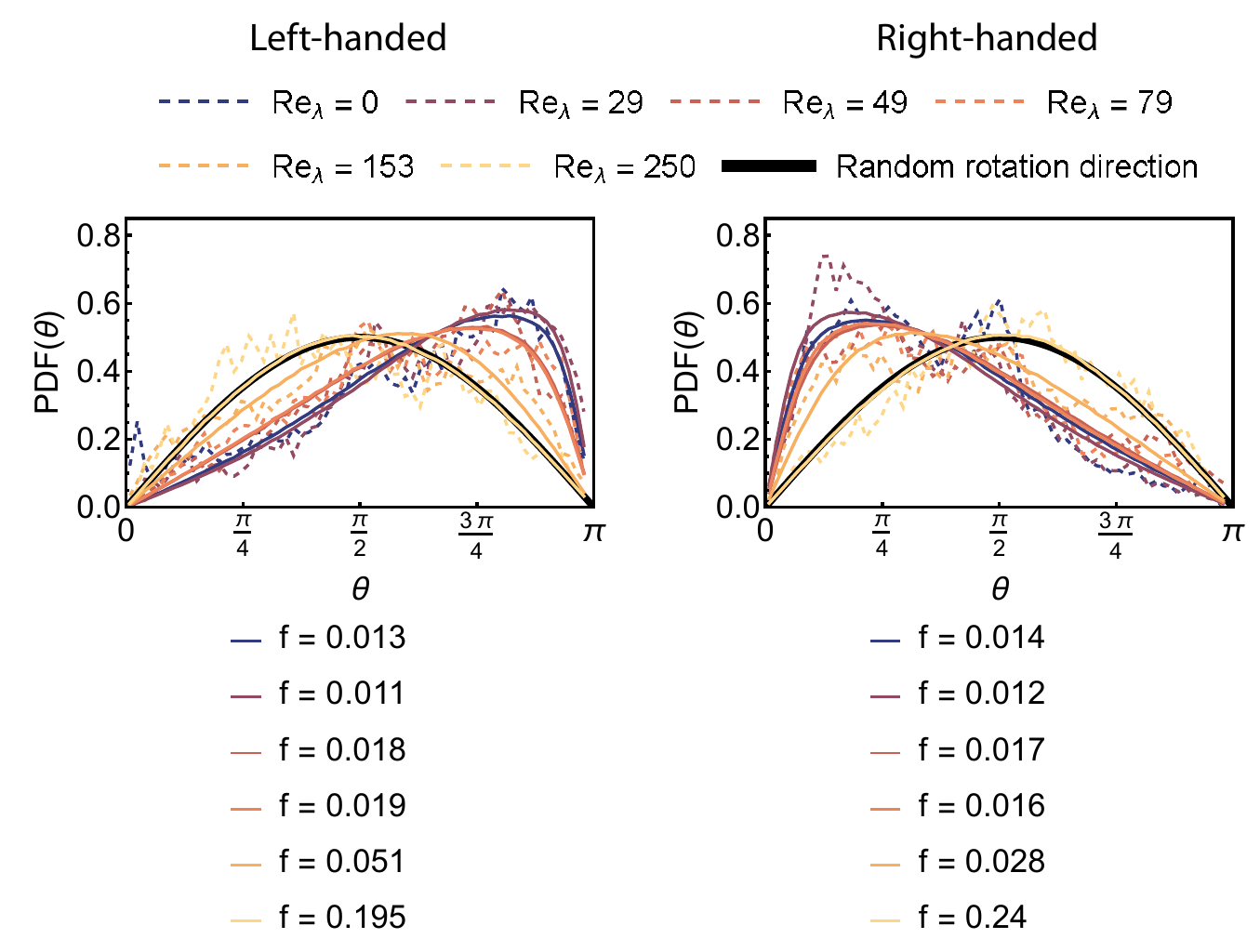}
    \caption{PDF of the angle between the rotation vector and the vertical for experiments (dashed lines), and theoretical model (solid lines). Data for left-handed particles shown in a), data for right-handed particles shown in b). The solid lines from theory are fitted to the PDFs from experimental data.}
    \label{fig:rotvecs_theory_pdf}
\end{figure}

Similarly, the PDF of the angle between the rotation axis and the vertical axis is computed using the theoretical model. Figure \ref{fig:rotvecs_theory_pdf} shows the PDFs for the model and experiments, where the values of stochastic forcing are again determined by fitting the theoretical curves to the experimental curves. Panel a) shows the data for left-handed chiral particles, whereas panel b) shows the data for right-handed chiral particles.
Once again, the theoretical model matches the experimental findings well, showing preferential rotation at small turbulence levels that vanishes at stronger turbulence. This shows that the theoretical framework of a simplified particle in Stokes flow with stochastic forcing can capture the most important trends of a particle settling under gravity in the presence of turbulence.

Some quantitative differences remain however: the magnitude of the stochastic force differs significantly between the orientation data and rotation data shown in Figures \ref{fig:upvecs_theory_pdf} and \ref{fig:rotvecs_theory_pdf}. Such differences are not unexpected given the heuristic nature of the turbulence model. Nonetheless, the theoretical model captures and clarifies the trends observed in experiments.\\
\\
The theoretical model reaffirms the interpretation of experiments, and highlights the transition in particle dynamics from stable to tumbling settling modes as a consequence of increasing turbulence. The corkscrew mode was surprisingly observed in experiments, where we found no conclusive cause of this specific settling mode. The theoretical model allows a close investigation of particle parameters, which may reveal the cause of the corkscrew mode.

\begin{figure}
    \centering
    \includegraphics[width=0.5\linewidth]{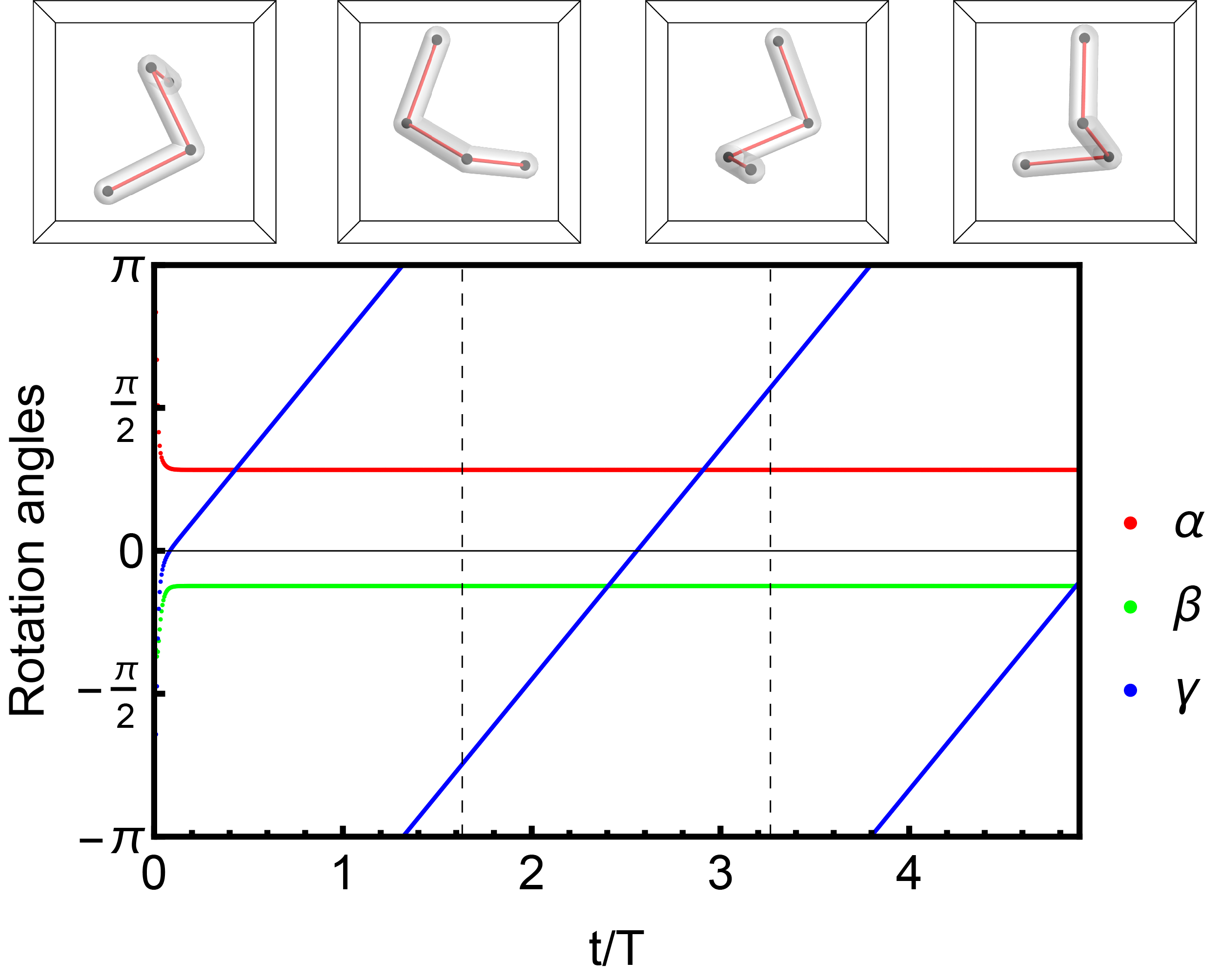}
    \caption{Theoretical model: Rotation angles for a left-handed chiral particle settling in Stokes flow. The snapshots illustrate the particle orientation at different time steps. In this case, one sphere is approximately $50\%$ lighter than the others. The lighter sphere is indicated in blue.}
    \label{fig:rotangs_corkscrew}
\end{figure}

Our experimental observations suggest that the attachment of a bubble causes the corkscrew settling mode. The effect of an attached bubble is implemented in the theoretical model by making one of the spheres lighter (by making it smaller), compared to the other spheres of the simplified chiral particle. The resulting settling dynamics of a chiral particle with an asymmetric mass distribution is shown in Figure \ref{fig:rotangs_corkscrew}, where one sphere's radius is $80\%$ as large as the others, resulting in a weight that is only $51\%$ of the other spheres. The lighter sphere is shown in blue.
This figure reveals the emergence of the corkscrew settling mode, which occurs for sufficiently large density imbalance. The rotation angles are similar to what was observed in experiments for the corkscrew settling mode, as shown in Figure \ref{fig:Rotangs_fallstyles}. Again we find a reversing of the rotation direction (around the vertical axis) compared to the stable settling mode. While the mass imbalance is very large in this example, it demonstrates that a mass imbalance can cause the occurrence of the corkscrew settling mode.

\begin{figure}
    \centering
    \includegraphics[width=0.5\linewidth]{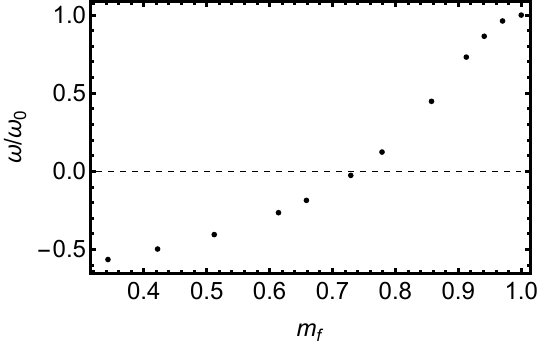}
    \caption{Theoretical model: Dimensionless angular velocity around the vertical axis as a function of the mass fraction of the single light sphere. The angular velocity is nondimensionalised by the angular velocity of the stable settling mode where all spheres have the same mass.}
    \label{fig:angvel_massfrac}
\end{figure}

This invites a more complete investigation of the mass imbalance using the theoretical model. We vary the mass of one of the spheres (at the end of the particle) as a fraction of the other spheres, and observe how the angular velocity around the vertical axis is affected. Figure \ref{fig:angvel_massfrac} displays the effect of the mass imbalance on the particle's angular velocity. The angular velocity is nondimensionalised by the angular velocity of a chiral particle consisting of spheres with equal mass (falling in a stable settling mode). The horizontal axis displays the mass fraction $m_f$, representing the mass of the lighter sphere as a fraction of the other spheres' mass.
The plot displays the observed rotation inversion, around $m_f = 0.73$. While this mass imbalance is too large to be caused by a bubble in experiments, this plot demonstrates that the mass imbalance in the particle (caused by a bubble, imperfections in the particle printing, or from another source) has significant effects on the dynamics of a settling chiral particle, and can cause a corkscrew settling mode. The supplementary material contains an animation of chiral particles with a single light spheres, which illustrates the effects of the mass imbalance, and shows the reversal of the rotation direction.

\section{Conclusion and Outlook} \label{sec:Conclusion and outlook}
We experimentally studied the settling of chiral particles for multiple turbulence intensities. The turbulence intensity is estimated using laser Doppler anemometry, giving the turbulence intensity range $0 \leq \text{Re}_\lambda \leq 250$. The measurements illustrate a decreasing falling (vertical) velocity and increasing horizontal root mean square velocity with increasing turbulence. In terms of orientation, the particle preferentially aligns opposite to the direction of gravity, and rotates around the vertical direction in a quiescent fluid. The rotation direction depends on the particle chirality, and both preferential orientation and rotation gradually vanish for increasing Taylor--Reynolds number $\text{Re}_\lambda$. This shows the existence of a stable settling mode, where the particle chirality dictates the rotation direction. As the turbulence increases, the effects of the particle chirality vanish.

At low Taylor--Reynolds numbers, the stable settling mode is frequently observed, whereas it disappears for higher Taylor--Reynolds numbers, where the chiral particles tumble. Surprisingly, the stable settling mode is still observed at moderately high Taylor--Reynolds numbers, illustrating that effects of particle geometry and chirality are still visible in moderate levels of turbulence. Besides these stable and tumbling settling modes, a `corkscrew' settling mode has been observed experimentally, which was not found in numerical simulations.

Our theoretical model confirms the experimentally observed trends, and indicates that a mass imbalance can cause the corkscrew settling mode. We find that bubbles attaching to the chiral particle indeed make the corkscrew mode more common, though we cannot conclude that these bubbles are the sole cause of the corkscrew mode, which calls for further investigations.


\backsection[Acknowledgements]{The authors wish to thank Nitay Ben Shachar for help with the theoretical model. We thank Giulia Piumini, Roberto Verzicco, Federico Toschi, and Xander de Wit for fruitful discussions on this topic. We wish to thank Duco van Buuren for help and discussions on laser Doppler anemometry and data analysis. The authors would also like to thank Gert-Wim Bruggert, Martin Bos, and Thomas Zijlstra for their technical support.}

\backsection[Funding]{This research was funded by the Dutch Research Council (NWO) under grant OCENW.GROOT.2019.031.}

\backsection[Declaration of interests]{The authors report no conflict of interest.}




\appendix

\section{Theoretical model}\label{app: theoretical model}

{The dynamics of a chiral particle are calculated using a bead-and-stick model in a Stokes flow. This replaces the true particle geometry with a minimal set of $N$ spheres that are chosen to capture the chiral particle's symmetry and its dominant fluid-structure physics. The concept of \cite{Krapf2009} is employed to calculate the resulting dynamics of the equivalent $N$-sphere particle, albeit using a different implementation which is detailed below. The imposed turbulence is included heuristically using stochastic forcing that is discussed in Section~\ref{turbsectiob}. The approach to choose the equivalent $N$-sphere particle is explained in Section~\ref{minimalspheres}.


Consider a general particle consisting of a set of $N$ spheres that are rigidly connected and immersed in a viscous fluid, so as to translate and rotate in unison. The settling dynamics of this general particle under gravity is calculated using standard hydrodynamics that employs the Oseen mobility tensor to account for the hydrodynamic coupling between the spheres~\citep{Kim1991}. This enables the velocity of each sphere, $\mathbf{v}_i$, where $i \in [1, N ] $, to be calculated provided spheres are well separated relative to their radii.

The mobility tensor, $\mathbf{M}_i^\text{single}$, for a single isolated sphere of radius, $a_i$, is given by
\begin{equation}
\mathbf{M}_i^\text{single} \equiv \frac{1}{6 \pi \mu a_i}  \mathbf{I},
\label{singlemobility}
\end{equation}
and connects the force, $\mathbf{f}_i$, applied to sphere $i$ and its resulting velocity, $\mathbf{v}_i$, via the relation,
\begin{equation}
\mathbf{v}_i = \mathbf{M}_i^\text{single} \cdot \mathbf{f}_i ,
\label{singlevelocity}
\end{equation}
where $\mathbf{I}$ is the identity tensor and $\mu$ is the fluid viscosity (set to unity in the code); Einstein summation convention is not used.

The Oseen mobility tensor, defined by
\begin{equation}
\mathbf{G}_{ij} \equiv \frac{1}{8 \pi \mu r_{ij}} \left( \mathbf{I} + \frac{\mathbf{r}_{ij} \mathbf{r}_{ij}}{r_{ij}^2} \right), \quad i \neq j,
\label{Oseentensor}
\end{equation}
gives the velocity $\mathbf{v}_i$ of sphere $i$ due to a force $\mathbf{f}_j$, applied to a different sphere $j$, by
\begin{equation}
\mathbf{v}_i = \mathbf{G}_{ij} \cdot \mathbf{f}_j , \quad i \neq j,
\label{Oseenveclocity}
\end{equation}
where $\mathbf{r}_{ij}$ is the position vector between the centers of spheres $i$ and $j$, and $r_{ij} \equiv |\mathbf{r}_{ij}|$ is the distance between them. The Oseen tensor provides the leading-order hydrodynamic coupling between the spheres, under the constraint that the sphere radii are much smaller than their separation, i.e., $a_i, a_j  \ll r_{ij}$.

The motion of a collection of $N$ spheres can then be calculated using the generalized mobility tensor, $\mathbf{M}$, which relates the total force vector, $\mathbf{f}$, to the total velocity vector, $\mathbf{v}$, of all spheres, 
\begin{equation}
\mathbf{v} = \mathbf{M} \cdot \mathbf{f},
\label{totalvelocity}
\end{equation}
where
\begin{equation}
\mathbf{f} \equiv [\mathbf{f}_1, \mathbf{f}_2, ..., \mathbf{f}_N]^T , \qquad \mathbf{v} \equiv [\mathbf{v}_1, \mathbf{v}_2, ..., \mathbf{v}_N]^T,
\label{netforcevelocity}
\end{equation}
are stacked column vectors composed of each applied force vector and sphere velocity vector, respectively. The generalized mobility tensor, $\mathbf{M}$, is thus a $3N \times 3N$ block matrix with $\mathbf{M}_i^\text{single}$ along its main diagonal and $\mathbf{G}_{ij}$  in the off-diagonal positions.

The generalized resistance tensor, $\mathbf{R}$, is the inverse of $\mathbf{M}$, i.e., $\mathbf{R} \equiv \mathbf{M}^{-1}$, which upon inverting Eq.~\eqref{totalvelocity}, gives the force applied to each sphere in terms of the sphere velocities,
\begin{equation}
\mathbf{f} = \mathbf{R} \cdot \mathbf{v}.
\label{totalforce}
\end{equation}


Consider a general particle consisting of $N$ rigidly-coupled spheres, due to translation and rotation about its center-of-mass, $\mathbf{r}_0$. The subsequent velocity of each sphere $i$ is
\begin{equation}
\mathbf{v}_i = \mathbf{u} + \mathbf{\Omega} \times \left(\mathbf{r}_i - \mathbf{r}_0\right),
\label{rigidbodyvelocity}
\end{equation}
where $\mathbf{u}$ is the translation velocity of the general particle and $\mathbf{\Omega}$ is the angular velocity about its center-of-mass. Equation~\eqref{rigidbodyvelocity} is then expressed in matrix form for compatibility with Eq.~\eqref{totalforce}, giving
\begin{equation}
[\mathbf{v}_1, \mathbf{v}_2, ..., \mathbf{v}_N]^T = \mathbf{V} \cdot [\mathbf{U}, \mathbf{\Omega}]^T ,
\label{rigidbodyvelocity1}
\end{equation}
where $\mathbf{V}$ is the $3N \times 6$ matrix,
\begin{equation}
\mathbf{V} =\left[\begin{array}{cc}\mathbf{I} & -\left(\mathbf{r}_1 - \mathbf{r}_0\right)_\times \\\mathbf{I} & -\left(\mathbf{r}_2 - \mathbf{r}_0\right)_\times \\... & ... \\\mathbf{I} & -\left(\mathbf{r}_N - \mathbf{r}_0\right)_\times\end{array}\right],
\label{skewsymmetricmatrix}
\end{equation}
with the skew-symmetric matrix representation for the cross-product, $\mathbf{a} \times \mathbf{b} \equiv \mathbf{a}_\times \cdot \mathbf{b}^T$, where
\begin{equation}
\mathbf{a}_\times \equiv \left[\begin{array}{ccc}0 & -a_3 & a_2 \\a_3 & 0 & -a_1 \\-a_2 & a_1 & 0\end{array}\right],
\label{skewsymmetricmatrixdefinition}
\end{equation}
for any vectors, $\mathbf{a} = [a_1, a_2, a_3]$ and $\mathbf{b} = [b_1, b_2, b_3]$. The minus sign in Eq.~\eqref{skewsymmetricmatrix} arises from use of the identity, $\mathbf{\Omega} \times \left(\mathbf{r}_i - \mathbf{r}_0\right) =  -\left(\mathbf{r}_i - \mathbf{r}_0\right) \times \mathbf{\Omega}$.

Substituting Eq.~\eqref{rigidbodyvelocity1} into Eq.~\eqref{totalforce}, and making use of Eq.~\eqref{netforcevelocity}, gives
\begin{equation}
\mathbf{f} = \mathbf{R} \cdot \mathbf{V} \cdot [\mathbf{U}, \mathbf{\Omega}]^T,
\label{totalforcerigid}
\end{equation}
which relates the total force vector experienced by all spheres to the translation and angular velocities of the general particle.

The net force, $\mathbf{F}$, experienced by the general particle is
\begin{equation}
\mathbf{F} \equiv \sum_{i=1}^N \mathbf{f}_i = [\mathbf{I}, \mathbf{I},..., \mathbf{I}] \cdot  \mathbf{f} ,
\label{totalforcegeneral}
\end{equation}
whereas the net torque, $\mathbf{\Lambda}$, about its center-of-mass is
\begin{equation}
\mathbf{\Lambda} \equiv \sum_{i=1}^N \left(\mathbf{r}_i - \mathbf{r}_0\right) \times \mathbf{f}_i  = 
\left[\begin{array}{cccc}\left(\mathbf{r}_1 - \mathbf{r}_0\right)_\times , & \left(\mathbf{r}_2 - \mathbf{r}_0\right)_\times , & ... & , \left(\mathbf{r}_N - \mathbf{r}_0\right)_\times\end{array}\right] \cdot \mathbf{f} \,.
\label{totaltorquegeneral}
\end{equation}
Collating Eqs.~\eqref{totalforcegeneral} and \eqref{totaltorquegeneral} then gives
\begin{equation}
[\mathbf{F}, \mathbf{\Lambda}]^T \equiv \left[\begin{array}{cccc}\mathbf{I} & \mathbf{I} & ... & \mathbf{I} \\\left(\mathbf{r}_1 - \mathbf{r}_0\right)_\times & \left(\mathbf{r}_2 - \mathbf{r}_0\right)_\times & ... & \left(\mathbf{r}_N - \mathbf{r}_0\right)_\times\end{array}\right] \cdot \mathbf{f} = \mathbf{V}^T  \cdot \mathbf{f} \, ,
\label{totalforcetorquegeneral}
\end{equation}
where we have used the definition of $\mathbf{V}$ in Eq.~\eqref{skewsymmetricmatrix}.

Finally, substituting Eq.~\eqref{totalforcerigid} into Eq.~\eqref{totalforcetorquegeneral} gives
\begin{equation}
[\mathbf{F}, \mathbf{\Lambda}]^T  = \mathbf{R}_\text{rigid} \cdot [\mathbf{U}, \mathbf{\Omega}]^T ,
\label{finalresistancequation}
\end{equation}
where the rigid-body resistance tensor is
\begin{equation}
\mathbf{R}_\text{rigid} \equiv \mathbf{V}^T  \cdot \mathbf{R} \cdot \mathbf{V}.
\label{rigidresistance}
\end{equation}

Equation~\eqref{finalresistancequation} is the result we seek that enables the translational velocity, $\mathbf{U}$, and angular velocity, $\mathbf{\Omega}$, of the general particle to be calculated for a given applied force, $\mathbf{F}$, and torque, $\mathbf{\Lambda}$. In the present study, motion of the general particle is calculated under constant gravity and torque free conditions, i.e.,
\begin{equation}
\mathbf{F} = -M_p g \, \hat{\mathbf{z}}, \qquad \mathbf{\Lambda} = \mathbf{0},
\label{bodyforce}
\end{equation}
where $M_p$ is the net particle mass, $g$ is the gravitational acceleration and $\hat{\mathbf{z}}$ is the Cartesian basis vector in the $z$-direction: the magnitude of the downward force is chosen to be unity in the implementation of this model. This is implemented by choosing an initial position and orientation for the general particle. The particle position and orientation are then time-stepped by solving Eq.~\eqref{finalresistancequation} for $\mathbf{U}$ and $\mathbf{\Omega}$, and updating the rigid-body resistance tensor, $\mathbf{R}_\text{rigid}$, at each time step. 

\subsection{Imposed turbulent flow}\label{turbsectiob}

To model the imposed turbulent flow, Gaussian stochastic noise with a mean of zero and a variance of $\sigma_\text{turb}$ is added to each Cartesian component of the calculated angular velocity at each time step. We do not concern ourselves with the translational velocity because this does not affect the resistance tensor in the laboratory frame, and hence the orientational dynamics for which we are primarily concerned. While this approach is heuristic, it provides a means by which the effects of turbulence can be assessed.

\subsection{Minimal set of spheres}\label{minimalspheres}
The true chiral particles measured in this study are represented in this theoretical model using a minimal set of spheres. The chosen sphere radii must be much smaller than all distances between them, as discussed above. The aim is to capture the dominant symmetries of the particle, and hence its fluid-structure physics, while not overly complexifying the model. This is achieved by  placing four spheres at the ends of each of the three straight rods; see Figure~\ref{fig:Chiral part theory}. The spheres are chosen to be of equal size for simplicity.

}

\bibliographystyle{jfm}
\bibliography{jfm}

\end{document}